\documentclass[showpacs, amssymb, amsfonts, amsmath,twocolumn,pre]{revtex4-1}

\usepackage{graphicx,epsfig}

\newcommand{\etal}{\textit{ et. al.}}

\begin{document}
\title{High-resolution dielectric study reveals pore size-dependent orientational order of a discotic liquid crystal confined in tubular nanopores}
\author{Sylwia~Ca{\l}us}
\affiliation{Faculty of Electrical Engineering, Czestochowa
University of Technology, 42-200 Czestochowa,
Poland}
\email[E-mail: ]{silvia.calus@gmail.com, andriy.kityk@univie.ac.at, patrick.huber@tuhh.de}
\author{Andriy~V.~Kityk}
\affiliation{Faculty of Electrical Engineering, Czestochowa
University of Technology, 42-200 Czestochowa,
Poland}
\email[E-mail: ]{silvia.calus@gmail.com, andriy.kityk@univie.ac.at, patrick.huber@tuhh.de}

\author{Lech~Borowik}
\affiliation{Faculty of Electrical Engineering, Czestochowa
University of Technology, 42-200 Czestochowa,
Poland}
\author{Ronan~Lefort}
\author{Denis~Morineau}
\affiliation{Institut de Physique de Rennes, UMR 6251, Universit\'e de Rennes 1, 35042 Rennes, France}
\author{Christina~Krause}
\author{Andreas~Sch\"onhals}
\affiliation{Federal Institute for Materials Research and Testing (BAM), D-12203 Berlin, Germany }
\author{Mark~Busch}
\author{Patrick~Huber}
\affiliation{Institute of Materials Physics and Technology, Hamburg University of Technology (TUHH), D-21073 Hamburg-Harburg, Germany
}
\email[E-mail: ]{silvia.calus@gmail.com, andriy.kityk@univie.ac.at, patrick.huber@tuhh.de}

\date{\today}

\begin{abstract}
We report a high-resolution dielectric study on a pyrene-based discotic liquid crystal (DLC) in the bulk state and confined in parallel tubular nanopores of monolithic silica and alumina membranes. The positive dielectric anisotropy of the DLC molecule at low frequencies (in the quasi-static case) allows us to explore the thermotropic collective orientational order. A face-on arrangement of the molecular discs on the pore walls and a corresponding radial arrangement of the molecules is found. In contrast to the bulk, the isotropic-to-columnar transition of the confined DLC is continuous, shifts with decreasing pore diameter to lower temperatures and exhibits a pronounced hysteresis between cooling and heating. These findings corroborate conclusions from previous neutron and X-ray scattering experiments as well as optical birefringence measurements. Our study also indicates that the relative simple dielectric technique presented here is a quite efficient method in order to study the thermotropic orientational order of DLC based nanocomposites.
\end{abstract}

\maketitle

\section{Introduction}
The self-organization of disc-like molecules can result in liquid crystalline phases with remarkable structural and dynamical properties \cite{Safinya1984, Fontes1988, Gharbia1992, Ehrentraut1995, Adam1993, Saunders2000, Bayer2003, DeCupere2006, Laschat2007, Kirkpatrick2007, Sergeyev2007, Elmahdy2008,Charlet2008, Mouthuy2008, Pouzet2009, Bisoyi2010, Hansen2011, Choudhury2011}. In particular, the  electrical conductivity and the high photoconductivity typical of columnar phases make discotic liquid crystals (DLCs) promising candidates for numerous technological applications, for example in electroluminescent and photovoltaic devices \cite{Schmidt-Mende2001, Wu2007, Demus1998, Kumar2014, Gentili2014}. 

\begin{figure*}[htbp]
\epsfig{file=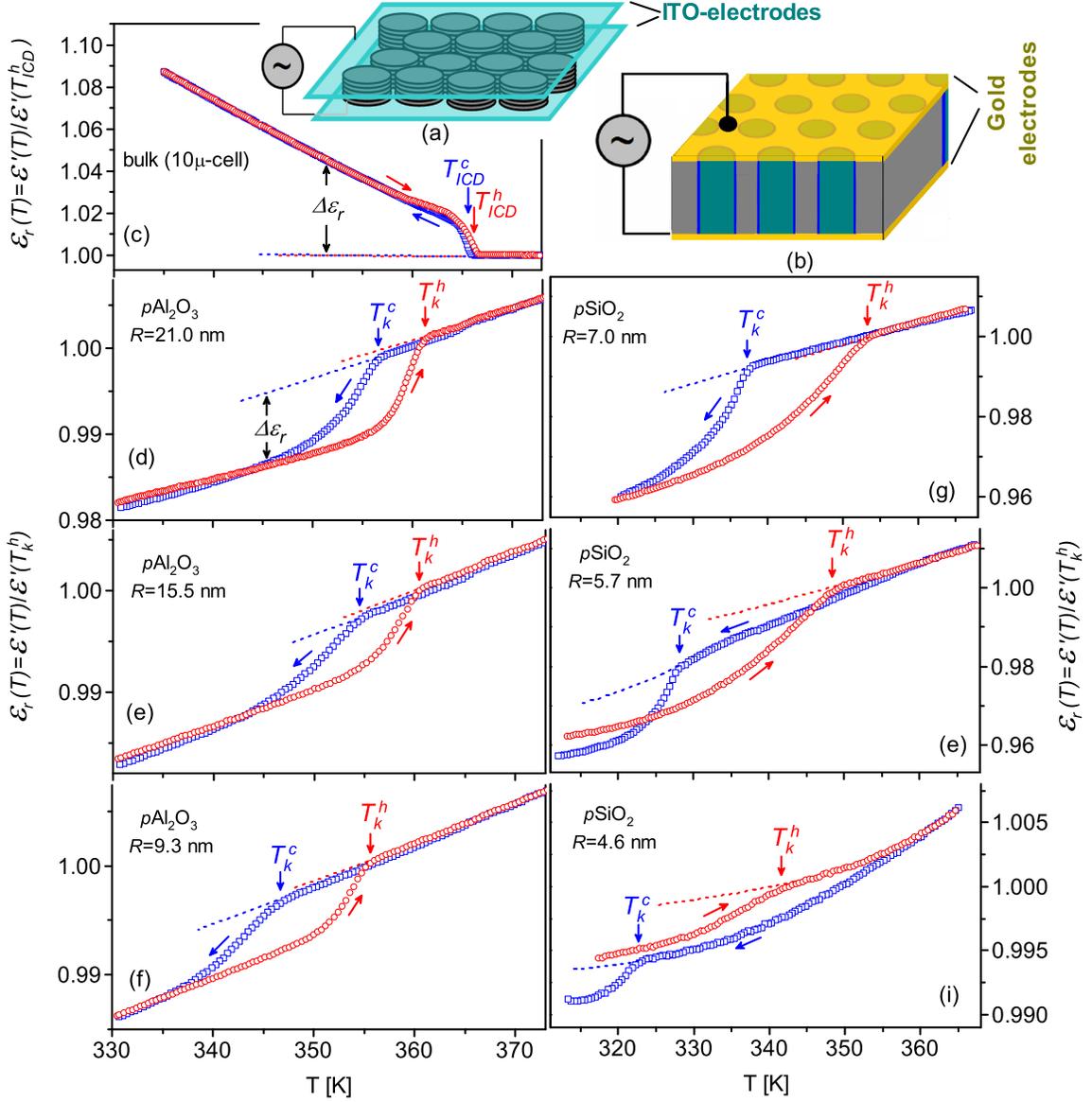, angle=0,width=1.8\columnwidth}\caption{(color online) Illustrations of the capacitor geometry used for the dielectric measurements in the bulk state (a) and for the nano-confined state (b). Reduced dielectric permittivity of Py4CEH vs. temperature during heating (circles) and subsequent cooling (squares) measured in the bulk state (c) and confined to mesoporous matrices of different pore radii of $p$Al$_2$O$_3$ (d,e,f) and $p$SiO$_2$ (g,h,i). The dashed lines are $\varepsilon_r(T)$-dependences extrapolated from the isotropic phase. $\Delta \varepsilon_r$  indicate the excess permittivity.} \label{fig1}
\end{figure*}

The molecular building blocks of many DLCs are polyaromatic cores surrounded by aliphatic chains. The stacking of the polyaromatic cores results in an hexagonal columnar structure. The one-dimensional conductivity originates in charge transfer along the molecular columns of the $\pi$-$\pi$ systems, whereas the alkyl side-chains insulate each column from its neighbors \cite{Bushby2005}. The columnar structure, therefore, represents an array of isolated conducting molecular wires. In comparison to conventional crystalline or amorphous conductors, liquid crystals have a number of advantages, such as simple processability, self-healing and spontaneous alignment. 

DLCs can also easily be embedded into nanoporous media by melt infiltration \cite{Zhang2009a,Corvazier2000, Duran2012, Kityk2014a, Zhang2014, Zhang2015, Huber2015}. One of the promising ideas in this respect is the synthesis of nanowires by this technique as pioneered by Steinhart et al \cite{Steinhart2005}. However, its practical implementation reveals difficulties to provide a long-range molecular order, which is easily maintained in the bulk state \cite{Charlet2008}. Possible strategies to overcome this problem are modifications of host-guest interactions, i.e. chemical modification of the pore wall along with a proper choice of the guest DLC material and the pore diameter \cite{Zhang2015}, or controlling the packing structure by applying external fields, e.g. by magnetic fields\cite{Takami2008}. 

However, similarly as the dynamical and structural behavior of rod-like liquid crystalline systems \cite{Crawford1991, Crawford1996, Kutnjak2003a, Kityk2008, Binder2008, Cordoyiannis2009,Kityk2010, Mazza2010, Chahine2010, Jasiurkowska2012, Karjalainen2013, Huber2013, Varga2014, Bono2015}, the properties of DLCs has turned out to be particularly susceptible to nano confinement \cite{Kopitzke2000,Steinhart2005, Stillings2008, Duran2012, Zhang2014, Kityk2014a, Calus2014,Zhang2015, Huber2015}. Thus, new, simple experimental methods that are able to monitor the type of molecular ordering inside nanoporous media are strongly demanded. Here, we demonstrate that a high-resolution dielectric study can yield important insights into the thermotropic, collective orientational order of a DLC and thus on confinement effects on its phase behavior. We compare those findings with results from experimentally more challenging and elaborated techniques, i.e. temperature-dependent neutron and X-ray diffraction \cite{Cerclier2012} as well as optical birefringence experiments \cite{Kityk2014a, Calus2014}.

The DLC used in our studies was the pyrene-1,3,6,8-tetracarboxylic rac-2-ethylhexyl ester (Py4CEH). It consists of a polyaromatic core surrounded by flexible aliphatic chains \cite{Hassheider2001}. In the bulk state Py4CEH forms a hexagonal columnar discotic (CD) phase (see Fig.~1a) between the isotropic (I) phase, which exists above 365~K, and the plastic crystalline phase observed below 246~K \cite{Cerclier2012, Krause2013}.

\section{Experimental}
We employed monolithic membranes of alumina ($p$Al$_2$O$_3$) and silica ($p$SiO$_2$), which are permeated by arrays of parallel, tubular channels. The $p$Al$_2$O$_3$ membranes (thickness $h$ =100 $\mu$m, Smart Membranes GmbH (Germany)) have been fabricated by electrochemical etching of pure aluminum in a sulphuric acid electrolyte. Their pore size has been determined by volumetric N$_2$-sorption isotherms recorded at $T = 77$ K which gives the following average pore radii, $R$: 21.0$\pm$2.0 nm (porosity P=24\%), 15.5$\pm$1.5 nm ($P$ =17\%) and 9.3$\pm$0.7 nm ($P$ =16\%) - see the top-view electron microgrograph of the membrane with 19 nm mean pore diameter in Fig.~2d. The $p$SiO$_2$ membranes have been fabricated by oxidation (12 h at $T$=800 $^o$C under standard atmosphere) of porous silicon obtained by electrochemical etching of boron doped silicon wafers (resistivity 0.01-0.02 ohm$\cdot$cm) using an electrolyte mixture of HF:C$_2$H$_5$OH (2:3) and a DC current density of 11-13 mA/cm$^2$ \cite{Lehmann2000, Sailor2011, Canham2015}. By varying the etching time and etching current, we have fabricated $p$SiO$_2$ membranes with pore radii 7.0$\pm$0.5 nm ($P$ = 52\%, $h$ =340 $\mu$m ), 5.6$\pm$0.4 nm  ($P$ = 35\%, $h$ =290 $\mu$m) and 4.6$\pm$0.3 nm ($P$ = 30\%, $h$ =180 $\mu$m).

Dielectric measurements have been performed with the Frequency Response Analyzer Solatron-1260A. For the bulk measurements a sample cell made of two parallel ITO-glass electrodes separated by a gap, $h$ = 10 $\mu$m, was used. The DLC has been aligned homeotropically, as sketched in Fig.~1a. 

A schematics of the capacitor design used in the dielectric measurements of the nanocomposites is shown in Fig.~1b. Gold electrodes have been evaporated onto the membranes. Before filling of the samples they were annealed for 15 minutes at $\sim$ 400~K in order to remove adsorbed pore water. Then the mesoporous membranes were completely filled by spontaneous imbibition (infiltration) in the isotropic phase at 370~K \cite{Gruener2011}. 

Note that in our capacitor geometry the channels are all arranged parallel to the applied electric field. The sample forms a simple parallel electrical circuit. Accordingly, the effective complex permittivity of the composite $\varepsilon^*$ is given by the permittivities of the host matrix $\varepsilon^*_m$ and the guest DLC $\varepsilon^*_d$  weighted with the their volume fractions: $\varepsilon^*=(1-P)\varepsilon^*_m+P\varepsilon^*_d$.  The temperature-dependent dielectric measurements have been performed with cooling and heating rates of 0.07~K/min.

The temperature changes of the host $p$Al$_2$O$_3$ or $p$SiO$_2$ matrices are weak and smooth in the entire temperature range with negligible losses in the frequency window 10$^2$ $\le f \le$ 10$^6$ Hz. Thus, the molecular ordering inside the pores caused by the isotropic-to-columnar hexagonal phase transition, which particularly influences the $\varepsilon^*_d$ permittivity, can be easily detected by analyzing the temperature dependence of the permittivity of the nanocomposite, $\varepsilon^*(T)$\cite{Grigoriadis2011}.

\section{Results and Discussion}
For the Py4CEH based nanocomposites the dielectric losses at frequencies below 1 kHz are due to conductivity related processes. These processes are not in the focus of this study. At higher frequencies, particularly close to the upper limit of the frequency window of the frequency response analyzer ($f \lesssim 10^6$ Hz) the dielectric losses of confined DLC Py4CEH are comparable to the experimental resolution used here \cite{Krause2012}. Because of the Kramers-Kronig relationship between dielectric loss and real part of the permittivity, the real part, $\varepsilon'(f)$, can be considered frequency-independent. This implies that the dipolar relaxation corresponds to the so-called quasi-static limit, $\tau \ll (2\pi f)^{-1}$, where $\tau$ is the corresponding relaxation time. This is in agreement with recent dielectric studies on bulk Py4CEH \cite{Krause2012, Krause2013}, where the Arrhenius plot (see Fig.~4 of Ref.\cite{Krause2012}) at the temperature region relevant here (from 310-370 K) give relaxation frequencies, $f_r=(2\pi\tau)^{-1}$, in the range between 30 and 500~MHz. Our measurements have been performed at a constant frequency of 200~kHz. This choice has the advantages that it results in a quasi-static character of the dielectric measurements ($f\ll f_r$) as well as an optimized accuracy and resolution, which depending on the measured capacitance changes from $\sim$ 10$^{-4}$ to 10$^{-3}$~pF (i.e. 0.003 to 0.01\% of the absolute values). As we will document below, this high resolution appears particularly crucial for our study, given the relatively weak changes of the dielectric permittivity associated with orientational ordering of the discotic molecules.


The absolute magnitude of the permittivity of the nano composite scales with the porosity and is, itself, not important for our analysis. Therefore, it is convenient to characterize the dielectric temperature behavior in the region of the phase transition by the reduced dielectric permittivities, $\varepsilon_r(T) =\varepsilon'(T)/\varepsilon'(T_{ICD}^h)$ (bulk state) or $\varepsilon_r(T)=\varepsilon'(T)/\varepsilon'(T_k^h)$ (confinement), expressed simply by the ratio of the measured capacitances $C'(T)/C(T_{ICD}^h)$ or $C'(T)/C(T_k^h)$, respectively. Here, $T_{ICD}^h$ and $T_k^h$ represent characteristic temperature points of the isotropic-to-columnar transition (discussed below) during heating in the bulk state and in confinement, respectively.
\begin{figure}[htbp]
\epsfig{file=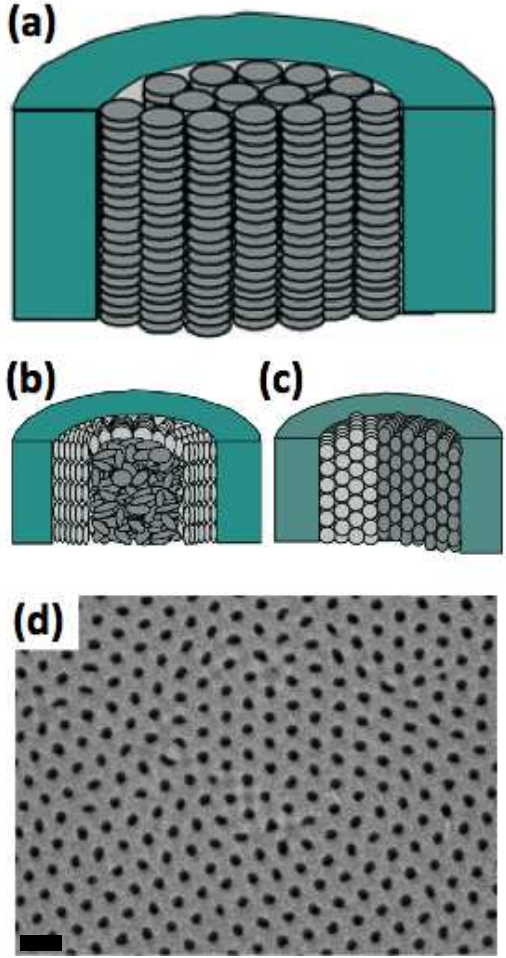, angle=0,width=0.8\columnwidth}
\caption{(color online) Molecular ordering types of DLCs in cylindrical nanochannels. (a) Parallel axial configuration. (b) Radial, face-on configuration with isotropic core. (c) Radial multidomain face-on configuration. (d) Top-view electronmicrograph of an alumina membrane with approx. 19 nm mean pore diameter. The scale bar indicates a width of 100 nm.}
\label{fig2}\end{figure}

In Fig.~1 we present the temperature dependences of the reduced dielectric permittivity, $\varepsilon_r(T)$, measured during heating and cooling runs for discotic Py4CEH in the bulk state and embedded into mesoporous matrices of different types and channel sizes.

Compared to the liquid crystalline phase, the bulk permittivity, $\varepsilon_r$, in the isotropic phase has only a weak temperature dependence. The isotropic-to-columnar transition is accompanied by a sudden rise of the reduced permittivity with a characteristic kink in the $\varepsilon_r(T)$-dependence at $T_{ICD}^c$, see Fig.~1c. Subsequent cooling reveals a small temperature hysteresis ($\sim$ 0.7 K) of the reverse transformation. In the bulk state the homeotropically aligned columnar phase exhibits a positive excess permittivity, $\Delta \varepsilon_r$. This may appear surprising at first glance, if one considers that at optical frequencies Py4CEH is characterized by a negative dielectric anisotropy ($\varepsilon_{\parallel, \infty}<\varepsilon_{\bot, \infty}$) which leads to a negative birefringence in the columnar discotic phase \cite{Kityk2014a, Calus2014}. It can, however, be rationalized as follows: The dielectric properties at optical frequencies are mainly determined by the \emph{electronic} polarizability. Here the contribution of the polyaromatic cores is dominant, which is characterized by a large polarizability anisotropy. At low frequencies, the \emph{dipolar} polarizability and the orientational polarization of molecular dipoles provide additional contributions to the dielectric permittivity. Accordingly, in the quasi static limit the dielectric permittivities are determined by $\varepsilon_{\parallel, st}=\varepsilon_{\parallel, \infty}+\Delta \varepsilon_{\parallel}$ and $\varepsilon_{\bot, st}=\varepsilon_{\bot, \infty}+\Delta \varepsilon_{\bot}$, where $\Delta \varepsilon_{\parallel}$ and $\Delta \varepsilon_{\bot}$ are the dipolar relaxation strengths. A positive dielectric anisotropy of Py4CEH DLC means that an inequality $\Delta \varepsilon_{\parallel}- \Delta \varepsilon_{\bot} > \varepsilon_{\bot, \infty}-\varepsilon_{\parallel, \infty}$ holds. Interestingly, the polyaromatic core does not contribute to the molecular dipole moment. Thus, the dipolar polarizability is attributable to the polar ester groups and the aliphatic chains forming the intercolumnar network in the hexagonal liquid crystalline structure. Their flexibility and ability to order as a function of temperature explain, presumably, why the quasi-static dielectric permittivity does not saturate at lower temperatures. This is in contrast to the saturation of the optical birefringence \cite{Kityk2014a,Calus2014} and thus to the saturation of the collective orientational ordering of the polyaromatic cores.

The temperature behavior of the reduced permittivity, $\varepsilon_r(T)$, for Py4CEH embedded into nanopores substantially differs from the bulk behavior as one can deduce from Figs.1d-1i. For all nanocomposites studied, the excess permittivity, defined as the difference between the measured value, $\varepsilon_r(T)$ and the one obtained by an extrapolation of the $\varepsilon_r(T)$-dependence from the isotropic phase (see marked by $\Delta \varepsilon_r$ in Fig.~1d for explanation), is found to be \emph{negative}. This means that the parallel axial columnar order (see sketch in Fig. 2a), which is particularly desirable for many nanotechnological applications, is not realized. The negative $\Delta \varepsilon_r$ is consistent with a molecular ordering for which discotic molecules are oriented preferably parallel to the long pore axis. The physico-chemical reason for the realization of such an ordering is the face-on type anchoring enforced by alumina and silica surfaces. 
It can lead to several different configurations, the most probable are the radial configuration with isotropic core (Fig.~2b) and the radial multidomain configuration (Fig.~2c) \cite{Cerclier2012, Kityk2014a, Zantop2015}. 

This observation is consistent with our X-ray and neutron diffraction study on Py4CEH in alumina and silica channels \cite{Cerclier2012}. It indicated that the molecular ordering is either of radial columnar or of radial nematic type for small channel diameter, see the X-ray and neutron diffraction patterns in Fig.~3 and Fig.~5, respectively, and the corresponding discussion in Ref. \cite{Cerclier2012}. 

\begin{figure}[htbp]
\epsfig{file=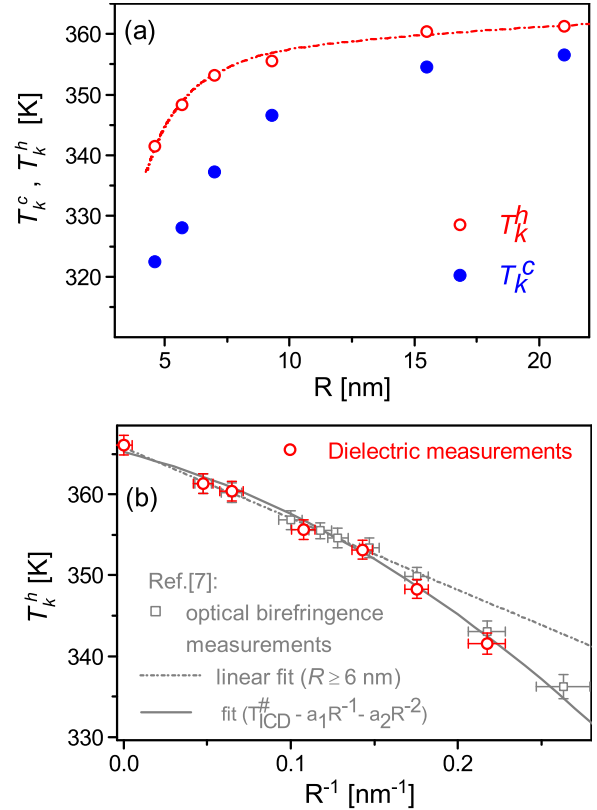, angle=0,width=0.9\columnwidth} \caption{(color online) (a): The characteristic temperatures $T^h_k$
and $T^c_k$ determined from the temperature dependences of the dielectric permittivity (see Fig.~1) measured during heating (open circles) and cooling (solid circles), respectively. The dash-dotted line is a guide for the eye. (b): $T^h_k$ vs. the inverse channel radius $R^{-1}$ as determined by dielectric measurements (open circles) and by optical birefringence data (open squares), respectively, in comparison with a $1/R$-fit (dash-dot line) and with a function $T^h_k=T^\#_{ICD}-a_1R^{-1}-a_2R^{-2}$ and the fit-parameters, $T^\#_{ICD}$=365.2 K, $a_1$=52.1 K$\cdot$nm and $a_2$=241.0 K$\cdot$nm$^2$ (solid line), see Ref.\cite{Kityk2014a} and discussion in the text.}
\label{fig3}
\end{figure}

As is obvious from Fig. 1 the geometrical constraints render the transition in the nanochannels gradual and hysteretic with characteristic lower and upper closure points (kinks) in the dielectric data at $T_k^c$ (cooling) and $T_k^h$ (heating). Just below these characteristic temperatures the excess permittivity, $\Delta \varepsilon_r(T)$,  exhibits strong variations with a saturation at lower temperatures. The larger the channel diameter, the steeper are the changes of $\Delta \varepsilon_r(T)$. The characteristic temperatures, $T_k^c$ and $T_k^h$, decrease with decreasing channel radius.

Note that, as discussed extensively in Refs. \cite{Kityk2014a,Calus2014} $T_k^c$ and $T_k^h$ do not represent phase transition temperatures in a classical meaning. The radial arrangement illustrated in Fig.~2b results in curved molecular layers, corresponding splay deformations and excess free energies upon collective molecular orientation. These excess energies are not homogeneous in pore space. Because of the increase in the curvature of the layers with decreasing distance $r$ from the channel centre, they decrease with $1/r$. Hence, the transition temperature towards the orientationally ordered state depends on the radial position of the layer. It decreases with decreasing $r$. During cooling the transition towards the radially ordered state starts at the periphery, near the channel walls, where the excess free energy is the lowest. Therefore, the interfacial interaction and the channel roughness play an important role for the onset of the phase transition. Radial homeotropic and axial planar ordering may simultaneously occur, enforced by specifics of the interfacial and/or local geometrical conditions. Accordingly, first a (metastable) coexistence of both states appear and only at lower temperature the radial molecular ordering becomes dominating. With these assumptions we could explain in our previous optical polarimetry study that the kink position upon cooling varied from sample to sample and was sensitive to the surface coating \cite{Kityk2014a}.

By contrast, the kink position during heating exhibited a more systematic evolution as a function of channel diameter and was independent of the chemistry of the channel surface. Presumably, there is always an entirely disordered or highly defect decorated core - see Fig. 2b. After all, the splay-deformation related excess energy diverges towards the pore center. This disordered core acts as a nucleation for the high-temperature phase. The phase front starts there and gradually propagates according to the $r$-dependent transition temperature to the periphery, while "melting" subsequent radially ordered molecular layers. The kink position at heating corresponds therefore to the situation at which the phase front reaches the channel wall. {It is defined mainly by a geometrical factor, i.e. the channel radius, and because of the omnipresent nucleation mechanism in the channel center it is a reliable measure to characterize the phase transition. For channels of radius $R$ with ideally smooth walls its temperature can be derived from a Landau-de Gennes free energy functional which includes the peculiar influence of radial-dependent excess energy outlined above \cite{Kityk2014a}, i.e.  $T_{\rm k}=T_{ICD}^{\#}-a_1R^{-1}-a_2R^{\rm -2}$. Here $T_{\rm ICD}^{\#}$=365.2~K is the bulk transition temperature, the $1/R$ term is analogous to the Gibbs-Thomson prediction for a phase transition of first order in cylindrical confinement \cite{Christenson2001, Alba-Simionesco2006, Berwanger2010} and the $1/R^2$ scaling results from the splay deformation contribution. 

In excellent agreement with these considerations also the $T$-shifts of the upper closer points of the hysteresis in the dielectric data, $T_k^h$ are dominated by two types of scalings, see Fig.~3: An $1/R$-scaling at large channel diameters, where the splay deformation energies are negligible, and an additional $1/R^2$-scaling for smaller channel diameters, which is characteristic of the excess energies related to splay distortions in the collective arrangement of the discotic molecules \cite{Kityk2014a}. Moreover, the prefactors of the scaling laws deduced from the $R$-dependence of the closure point shifts agree with the ones derived in the birefringence study, i.e.  $a_1=$ 52.1 K$\cdot$nm and $a_2=$ 241.0 K$\cdot$nm$^2$.}

\section{Summary}
In summary, we have reported a high-resolution dielectric study on the columnar DLC Py4CEH confined in tubular nanopores. In contrast to the optical birefringence, the dielectric permittivity of Py4CEH in the quasi-static limit is characterized by a positive dielectric anisotropy in the columnar discotic phase. This determines the dielectric behavior in the region of the isotropic-to-columnar discotic transition both in the bulk and in the confined states. Whereas the homeotropic alignment in the bulk state is characterized by a positive excess permittivity, molecular ordering of Py4CEH in the confined cylindrical geometry is accompanied by a negative value. It corresponds to ordered structures, where discotic molecules orient preferably parallel to the long channel axis (in a so-called face-on configuration). The $R-$scaling of characteristic phase transition temperature points, deduced from the dielectric data, are in very good agreement with recent neutron scattering and optical birefringence studies on the same discotic mesogen. 

Molecular dynamics simulations on rod-like systems have successfully contributed to a more detailed understanding of the isotropic-to-nematic transition in confinement \cite{Gruhn1998, Binder2008, Ji2009, Ji2009b, Mazza2010}. In fact, a recent molecular dynamics study documented radial movements of phase fronts analogous to the ones inferred here, however for a rod-like system in nanochannels\cite{Karjalainen2013}. Molecular dynamics simulations on confined and bulk DLC \cite{Busselez2014, Zantop2015} indicate, in agreement with the experiments presented here, that the phase behavior of the confined system strongly differs from its bulk counterpart: the bulk isotropic-to-columnar transition is replaced by a continuous ordering from a paranematic to a columnar phase. Moreover, a reorganization of the intercolumnar order was observed which reflects the competing effects of pore surface interaction and genuine hexagonal packing of the columns.

This competition between surface anchoring and tendency to form rigid, hexagonal columnar phases has turned out to be decisive for the formation of either axial or radial or circular concentric arrangements of the columns in cylindrical nano confinement \cite{Kopitzke2000,Steinhart2005, Stillings2008, Duran2012, Zhang2014, Kityk2014a, Calus2014,Zhang2015, Huber2015}. For edge-on anchoring Zhang \etal \cite{Zhang2014, Zhang2015} found a transition from circular concentric towards an axial columnar arrangement upon increase in columnar rigidity of a triphenylene based DLC. This result corroborates the observations of 
Duran \etal \cite{Duran2012}. They found for a DLC with a comparably large core (hexabenzocoronene), and thus presumably large column rigidity, confined in tubular alumina nanopores an axial arrangement of the columns in the channel center, despite a face-on anchoring at the pore walls. Moreover, below a certain pore diameter the transition towards a low-$T$ crystalline phase, typical of the bulk of this system, is completely suppressed. 

Diffraction experiments and Molecular Dynamics simulations on partial fillings \cite{Huber2013, Calus2015} may allow for complementary structural information on the propagation of the phase boundaries in channel space. This will shed light on the remarkably different pathways found here for freezing and melting of discotics.

\begin{acknowledgments}
This work was supported by the Polish National Science Centre (NCN) under the project ''Molecular Structure and Dynamics of Liquid Crystals Based Nanocomposites''(Decision no. DEC-2012/05/B/ST3/02782) and by the French-German Project TEMPLDISCO jointly founded by ANR (ANR-09-BLAN-0419) and the German Science foundation (DFG SCHO 470/21 and HU 850/3). We are grateful to Julien Kelber and Harald Bock (CRPP) for having resynthesized and provided the liquid crystalline material studied here.
\end{acknowledgments}
%

\end{document}